# Ownership in the Hands of Accountability at Brightsquid

A Case Study and a Developer Survey

Umme Ayman Koana\*, Francis Chew˙, Chris Carlson˙, Maleknaz Nayebi\*

\*EXINES Lab, York University, ˙Brighsquid

{koana,mnayebi}@yorku.ca,{francis,Chris.Carlson}@Brightsquid.com

## ABSTRACT

The COVID-19 pandemic has accelerated the adoption of digital health solutions. This has presented significant challenges for software development teams to swiftly adjust to the market need and demand. To address these challenges, product management teams have had to adapt their approach to software development, reshaping their processes to meet the demands of the pandemic. Brighsquid implemented a new task assignment process aimed at enhancing developer accountability toward the customer. To assess the impact of this change on code ownership, we conducted a code change analysis. Additionally, we surveyed 67 developers to investigate the relationship between accountability and ownership more broadly. The findings of our case study indicate that the revised assignment model not only increased the perceived sense of accountability within the production team but also improved code resilience against ownership changes. Moreover, the survey results revealed that a majority of the participating developers (67.5%) associated perceived accountability with artifact ownership.

## KEYWORDS

Ownership, Accountability, Software Quality, Software Engineering

## 1 INTRODUCTION

The emergence of the COVID-19 global pandemic has accelerated the transformation of digital health solutions, as the increasing demand for online appointments and remote healthcare services became more pressing due to social distancing guidelines. This sudden shift has resulted in significant changes in the digital health industry and has added new challenges for software development teams, particularly in the healthcare sector. Former studies showed that the COVID-19 pandemic has had a profound impact on software development teams [7, 15, 30, 40]. In particular, the surge in demand for digital health solutions has created a new set of challenges for software teams, including shifting priorities, increased workload, and rapidly changing customer demands. In response to these challenges, product management teams have had to adapt and evolve their approach to software development. To overcome the challenges posed by the pandemic, software development teams have had to change the shape and format of their product management processes.

Brightsquid[1] is specialized in providing secure communication solutions for the healthcare industry. The company's mission is to help healthcare providers communicate more effectively and securely, enabling them to improve patient outcomes and deliver better quality care. Brightsquid's flagship product, the Secure-Mail platform, is HIPPA[2] compliant and offers secure messaging and file-sharing capabilities, enabling healthcare professionals to exchange protected health information in a safe and efficient manner [31, 32]. Brightsquid has been involved in several projects and has experienced an increasing demand for new or enhanced versions of their existing solutions during the COVID-19 pandemic. The company's product management team has made changes to their issue assignment model to increase developers' accountability toward end-user, responding to the increased market demand.

The company explains this change as the transition from task assignment to design assignment. In this new approach, a user story is assigned to developers, and it is their responsibility to identify relevant tasks and expert team members. While the team perceived an increase in developers' accountability, we initiated the partnership to evaluate the impact of this change on code ownership and code quality, and to what extent it affects them. We started with an interview with the product manager and the Chief Technology Officer (CTO) of Brighsquid to identify the problem.

### 1.1 Developer Accountability in the Face of High Customer Demand

With the rise in demand during the COVID-19 pandemic, the company recognized the critical importance of enhancing development team accountability toward customers. The team leads and product managers have increasingly noticed that developers tend to limit their contributions to simply carrying out assigned tasks, as suggested by the leads. The company has identified the significance of fostering a culture of co-creation to promote enhanced product innovation and increase customer satisfaction. As a result, the team opted for a change that holds developers further accountable for the outcomes of an implementation task:

> "We intended to build a self-organizing team, which resembles how jazz musicians organize themselves. We have a broad structure and a shared destination that everyone understands and should commit to. Despite not having been together (during COVID), we have embraced this approach of self-organization. With the intent to bring more knowledge, curiosity, diligence, code ownership, and code robustness to the development team Brightsquid gradually changed the task assignment process to be more flexible in the last six months."

To achieve this vision, the team leads emphasized a shift from task assignment and ownership to user story assignment and design ownership. Figure 1 - (a) and Figure 1- (b) illustrate the process before and after the assignment shift. Before the shift in the task assignment process, the company followed a standard approach where the product manager identified the user story and collaborated with the users, market team, and customer management team to determine the specifics of the user story. The product manager then identified the tasks required to implement that user story. Subsequently, in a collaborative effort with the developers, the tasks were prioritized and assigned to developers in a sprint. The team has been practicing agile practices such as self-assignment following the agile practices [23].

---
[1]https://Brightsquid.com/
[2]HIPAA: Health Insurance Portability and Accountability Act of 1996

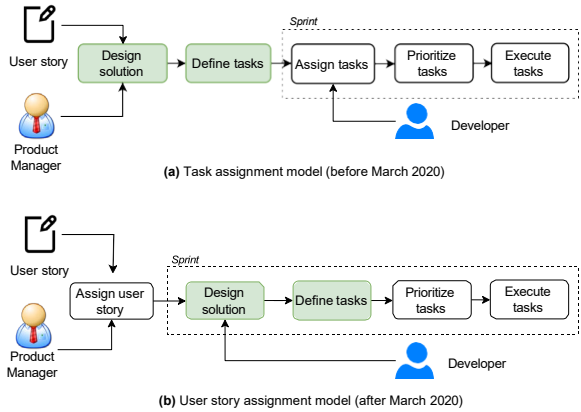

**Figure 1: The process of (a) product manager details tasks for every user scenario and assign to a developer - Before March 2020 (b) product manager assign user story to a developer who is then responsible to design solution and define tasks.**

However, after the change (Figure 1 - (b)) in the process, once the product manager identifies and details the user stories, they then identify the most suitable developer based on expertise and assign the story to them. From there, the developer takes on the responsibility of making decisions regarding design and implementing solutions. They identify the necessary tasks, enter them into the issue-tracking system, and prioritize them. Additionally, they collaborate with other team members to complete the assigned tasks. The product manager remains involved in the process to provide insights from customers, the customer management team, and the business perspective. In this scenario, the developer assigned to the user story also assumes accountability for finalizing and reporting on the story's progress. The company perceived a general value in this change, as put by Brightsquid's CTO:

*"It is the developers' responsibility to get people together to talk about the design solutions or add subtasks under that user story based on their solution. (By this change) we observed greater collaboration and participation. Developers were no longer just being order takers but also participating actively in figuring out what ought to be done."*

While the process aligns with the established transition of management in agile methodologies [24], the organization had only a primitive understanding and description of developers' accountability and how to measure it. As framed by the product manager:

*"Generally the idea was that whoever is assigned accountability for a particular story, if it succeeds, it reflects positively on them, but if it fails, it's like, 'Hey, you didn't fulfill your responsibilities'."*

In this context and as a result of the semi-structured interview and its comparison with the literature, we formulate three research questions to evaluate the impact of this change on code ownership. To further gain a deeper understanding of the significance of accountability among developers. We aimed to explore a more precise definition of accountability in software teams and examine its relationship with code ownership among software developers.

### 1.2 Research Questions

Specifically, this paper is focused on answering the following research questions:

**RQ1:** How has the relationship between ownership and software quality metrics changed following the implementation of the new issue assignment model at Brightsquid?

While Brightsquid aimed to foster a sense of accountability among developers, the impact of this change on the quality of the product was unclear. Therefore, we are interested in comparing the code ownership status of Brightsquid's project and the impact it has on the quality of the code before and after the task assignment change. We chose state-of-the-art studies and replicated them within BrightSsquid to measure the code ownership status in relation to code quality before and after the process change in March 2020.

**RQ2:** How did the revision of BrightSquid's issue assignment process affect the performance of prediction models in identifying defective files and directories?

We aim to investigate the efficacy of ownership metrics in developing a classification model for predicting defective files and directories in Brightsquid projects. To assess the impact of the new assignment model implemented in March 2020, we compare the performance of these models before and after the process change, evaluating whether the new model improves the predictability of defects. Furthermore, we analyze the significance of ownership metrics for both files and directories to determine the extent to which different metrics can predict code business before and after the process change.

**RQ3:** How do developers perceive the relationship between accountability and ownership?

The relationship between accountability and ownership is a complex and multifaceted issue that is perceived differently by developers depending on a variety of factors. We surveyed 67 participants to explore their interpretation of accountability and ownership and the relation between these two as they perceive.

## 2 BRIGHSQUID DATA

Brightsquid has successfully completed 39 projects with the collaboration of 52 developers between 2012 and 2023, each project commencing at different times. The company employs GitHub as a repository for storing its source code and leverages JIRA, an effective issue-tracking system, to efficiently manage its diverse range of issues. Brighsquid practice proper tracing method between code and issue repository.As of the time of writing this paper, Brightsquid had amassed an impressive count of 10,110 JIRA tickets across eight distinct categories: Story, Document, Task, Bug, New Feature, Epic, Technical Debt, and Improvement. These categories consist of 180 Epics, 1,243 Stories, 5,170 Bugs, 1,727 Improvements, 1,135 Tasks, 34 New Features, 548 Documentation, and 73 Technical Debts. To ensure a structured approach, the Story and Task issues are further classified into sub-stories, sub-tasks, and technical tasks, amplifying the organization's ability to address complex problems. Meanwhile, Bug, Epic, and Improvement issues are divided into sub-stories and technical tasks, enabling a systematic resolution of these concerns. Notably, Technical Debt, New Features, and Documentation issues do not possess any subdivisions, streamlining their management. Figure 2 provides an overview of this structure.

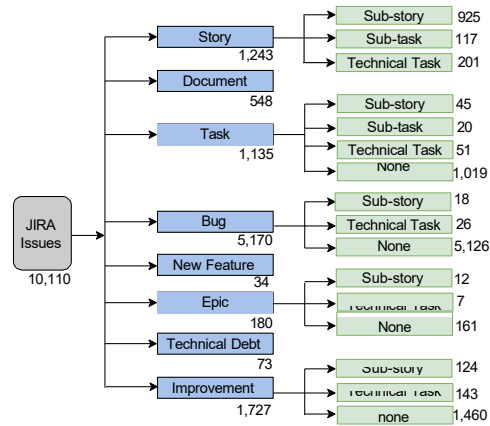

**Figure 2: JIRA issue hierarchy of Brightsquid**

**Table 1: Characteristics of the studied Brightsquid projects.**

| ID | Commits | Files | Changed files | Bugs |
|---|---|---|---|---|
| *March 2018 to March 2020* | | | | |
| Proj1 | 125 | 75 | 20 | 200 |
| Proj2 | 131 | 60 | 48 | 64 |
| Proj3 | 768 | 96 | 177 | 344 |
| Proj4 | 1,122 | 312 | 118 | 640 |
| *March 2020 to March 2022* | | | | |
| Proj1 | 53 | 34 | 11 | 10 |
| Proj2 | 82 | 14 | 9 | 10 |
| Proj3 | 317 | 96 | 53 | 147 |
| Proj4 | 228 | 282 | 65 | 100 |

For our paper, we focused on four significant projects from Brightsquid that remained highly active between 2018 and 2022. These four projects contain a total of 1,515 bug fixes. We collected data for two distinct time frames: March 2018 to March 2020, prior to the implementation of the new accountability system, and March 2020 to March 2022, following the change. During our considered time frame from March 2018 to March 2020, Proj4 has the most commits (1,350) and bugfixes (740) while Proj2 fixed the least bugs (74). On the other hand, Proj1 and Proj3 have 178 and 1,085 commits and 210 and 491 bugfixes respectively. For these four projects Table 1 shows the summary of these projects indicating the number of commits, total files, and files that have been changed within the time period, bug fixes, and number of developers for two distinct time periods.

## 3 BACKGROUND

Software engineering has a significant body of research on code ownership [39, 44, 45], and Nordberg et al. [37]. For our research, we adopted the methodologies employed in two prominent studies that examined ownership status before and after March 2020. Bird et al. [9] conducted a comprehensive study at Microsoft, later replicated by Greiler et al. [18], which focused on ownership metrics and their impact on software quality. We selected these studies as they were conducted within a similar context to ours at Brightsquid, involving proprietary products.

Bird et al. [8] introduced four ownership metrics at the file level: Major, Minor, Ownership, and Total, which are further explained in Table 2. They investigated the relationship between ownership metrics and pre- and post-release bugs for Windows Vista and Windows 7. In their study on ownership, Bird et al. [9] demonstrated that software components with numerous Minor contributors tend to experience more failures compared to components with fewer contributors. Conversely, components with a high level of Ownership tend to have fewer failures than those with lower ownership levels. The researchers also found that contributors classified as Minor for one component often become Major contributors for other components that share dependency relationships. Furthermore, their study revealed that removing Minor contribution information from defect prediction techniques significantly reduces their performance. Greiler et al.[18] replicated the former study by Bird et al.[9] and expanded upon it by introducing 11 ownership metrics at the directory level. They developed a classification model to predict bugs and assessed the bug-predictive power of the ownership metrics for both files and directories. Additionally, Greiler et al. adjusted the threshold for identifying Modest ownership, considering ownership below 50% as the criterion. They also introduced a new concept called Minimal ownership, which referred to developers with less than 20% ownership. We provided a summary of all the ownership metrics in Table 2 and we calculated each metric for every file or directory at the end of the designated time period.

## 4 EMPIRICAL METHODOLOGY

The primary objective of our study is to compare the ownership status and metrics before and after Brightsquid transitioned from assigning tasks to assigning user stories to software developers. This shift in the process serves as a key focus of our analysis, allowing us to evaluate the impact on ownership and associated metrics.

**Table 2: Definition of metrics inferred from literature and used in our study.**

| Metric | Definition |
|---|---|
| Major | # of developers who committed more than 5% to a file |
| Minors | # of developers who committed less than 5% to a file |
| Minimals | # of developers who made less than 20% commits |
| Modests | # of developers who made less than 50% commits |
| Total | # of developers who contributed to a file |
| Ownership | Proportion of commits for the highest contributor to a file |
| Avgownership | Average ownership values for all files in a directory |
| Ownershipdir | % of commits of the highest contributor among all files in a directory |
| Minownerdir | % of commits of the lowest contributor among all files in a directory |
| Avgcontributors | Average of distinct contributors among all files in a directory |
| Pcminors | % of contributors among all contributors with less than 50% commits for all directory files |
| Pcminimals | % of contributors among all contributors with less than 20% commits for all directory files |
| Pcmajors | % of contributors among all contributors with more than or 50% commits across all directory files |
| Avgminimal | Average minimals in a directory |
| Avgminors | Average minors in a directory |
| Minownedfile | The ownership value of the file with the lowest ownership value |
| Weakowneds | # of files in a directory that have an ownership value of less than 50% |

## 4.1 The relationship between code ownership and code quality metrics (RQ1)

We start by calculating the Spearman rank correlations between code ownership metrics and the number of bugs across the four projects. Our objective is to investigate the relationship between ownership metrics and software quality. We examined these correlations at both the file and directory levels. We explored the association between the number of bugs and four ownership metrics: Ownership, Total, Major, and Minor. We constructed a multiple linear regression model to examine the relationship between ownership metrics and the number of bugs. This model enables us to determine whether the increase in the number of bugs within projects can be attributed to code ownership metrics (for example, higher involvement of minor contributors) or to code-related factors (for example code size, churn, and complexity), which are known to be linked to software faults. By utilizing multiple linear regression, we analyzed the association between ownership metrics and bugs while accounting for source code attributes, specifically size, churn, and complexity. For each project, we created five statistical models to assess the extent of the impact that each metric has on the number of bugs. Initially, we established the "Base" model, which incorporated the three code metrics size, churn, and complexity. Subsequently, we iteratively add each ownership metric individually to identify the metrics that exerted an influence. We compare these models in terms of their variance in bugs ($R^2$).

For both calculating the correlation and constructing the linear regression models, we divided the data into two segments before and after March 2020. This division was made to account for the change in the assignment model implemented by Brightsquid. We then compared the correlation and the variance in bug measures for each project and across all the projects. Also, we investigated **RQ1** at both the file level and directory level. Bird et al. [9] conducted an analysis of `Microsoft Windows` products using file-level binaries. Following that, Greiler et al. [18] conducted a study that examined ownership in `Windows` products at the directory level. Their results showed an enhanced performance of defect prediction models when ownership was evaluated at the directory level.

## 4.2 Bug predictability performance (RQ2)

To investigate RQ2, we initially utilized a Random Forest classifier to predict defective source files and directories in Brightsquid. This choice of approach was influenced by the methodology employed by Greiler et al. [18] when analyzing ownership in `Microsoft`. The performance of our models was evaluated using 10-fold cross-validation. To assess the impact of the assignment process change on defect predictability, we constructed separate models for two distinct time periods: the two years prior to March 2020 and the two years following it. This division enables us to examine whether the predictability of defects improved (or not) after the assignment process change. Furthermore, we conducted training and testing of our models at both the file level and the directory level to provide a comprehensive analysis.

To conduct a more comprehensive analysis and evaluate the prediction power for each metric, we performed a metric importance analysis to evaluate the predictive capability of different metrics at the file and directory levels. In this evaluation, we specifically focused on the files and directories that underwent changes. The analysis was carried out using the Random Forest Regressor model from the `scikit-learn` library in `Python`. We then compared the predictive power of these metrics before and after March 2020, when Brightsquid changed its assignment model. This analysis allows us to assess the impact of the assignment model change on the effectiveness of the metrics in predicting defects.

## 4.3 Developer's perception of ownership and accountability in software teams (RQ3)

we discussed the results of **RQ1** and **RQ2** and their implications with the Brightsquid team. In order to gain a broader understanding of developers' perceptions in general, beyond just Brightsquid, and to investigate the relationship between code ownership and accountability in software teams, we have performed survey research following the guidelines outlined by Pfleeger and Kitchenham[38]. Our survey research consists of four major parts. The first part aimed to collect participants' demographics. The second part aimed to investigate the source and degree of a developer's sense of ownership within a team then we aimed to understand the definition of accountability at work for software developers. Lastly, we investigated the relationship between ownership and accountability. We designed the survey together with the Brightsquid team to evaluate their hypothesis in a broader population.

In total, the survey consisted of 26 questions, including 24 closed-ended questions and two open-ended questions. We collected demographic information through four questions. The remaining questions aimed to gather participants' opinions, experiences, and decisions using a five-point Likert scale, multiple-choice options, and text boxes. We obtained ethics approval for this survey from the York University Board of Ethics with Certificate # : $e2023 - 088$. The survey was anonymous, and no personal information was collected from participants. We used `Qualtrics` as the survey instrument. We used convenient sampling for attracting participants [38] and distribute the survey in our social networks. Overall, 128 times the link was clicked and 67 developers participate in our survey.

**Table 3: Spearman correlation between metrics and bug numbers on file level.**

| | March 2018 to March 2020 | | | | |
|---|---|---|---|---|---|
| **Ownership Metrics** | **Proj1** | **Proj2** | **Proj3** | **Proj4** | **Avg.** |
| Ownership | -0.217 | -0.214 | -0.394 | -0.432 | -0.314 |
| Major (>5%) | 0.236 | 0.259 | 0.43 | 0.548 | 0.368 |
| Minors (<5%) | 0.236 | 0.213 | 0.393 | 0.429 | 0.317 |
| Modests (<50%) | 0.236 | 0.214 | 0.40 | 0.43 | 0.320 |
| Minimals (<20%) | 0.217 | 0.214 | 0.184 | 0.425 | 0.26 |
| Total(NumDevs) | 0.236 | 0.237 | 0.412 | 0.50 | 0.346 |
| **Code Metrics** | | | | | |
| Churn | 0.211 | 0.233 | 0.416 | 0.539 | 0.350 |
| Size | 0.438 | 0.191 | 0.464 | 0.293 | 0.346 |
| Complexity | 0.217 | 0.214 | 0.184 | 0.425 | 0.26 |
| | March 2020 to March 2022 | | | | |
| **Ownership Metrics** | **Proj1** | **Proj2** | **Proj3** | **Proj4** | **Avg.** |
| Ownership | -0.334 | -0.261 | -0.422 | -0.457 | -0.369 |
| Major (>5%) | 0.349 | 0.266 | 0.433 | 0.474 | 0.380 |
| Minors (<5%) | 0.334 | 0.261 | 0.422 | 0.465 | 0.370 |
| Modests (<50%) | 0.296 | 0.235 | 0.192 | 0.102 | 0.206 |
| Minimals (<20%) | 0.334 | 0.261 | 0.174 | 0.317 | 0.271 |
| Total (NumDevs) | 0.343 | 0.264 | 0.428 | 0.465 | 0.375 |
| **Code Metrics** | | | | | |
| Churn | 0.434 | 0.265 | 0.428 | 0.477 | 0.401 |
| Size | 0.392 | 0.365 | 0.093 | 0.339 | 0.297 |
| Complexity | 0.334 | 0.261 | 0.174 | 0.317 | 0.271 |

## 5 RESULTS

In this section, we respond to each research question in sequence.

### 5.1 Relationship between code ownership and code quality metrics (RQ1)

We calculate the correlation between ownership metrics (see Table 2) and the number of bugs for each project at Brightsquid. We did this analysis for files and for directories. We did this separately for the time period of March 2018 to 2020 and the period of March 2020 to March 2022. We first present file-level results and follow the directory-level evaluations.

At the file level, Table 3 presents a comparison of the Spearman correlations between ownership metrics and the number of bugs at the file level before and after March 2020, when the assignment model changed in Brighsquid. In both time periods, we observed a negative correlation between the metric "ownership" and the number of bugs in the projects consistently across all projects. This indicates that higher levels of code ownership are associated with a decrease in the number of bugs. Furthermore, we observed a stronger negative correlation between bugs and ownership for all the projects after the assignment model changed (after March 2020). Additionally, we found that contributors classified as "Minor," "Minimal," and "Modest" all showed a positive correlation with the number of bugs in the code across all projects. However, none of the correlations are strong (all are < 0.5).

Furthermore, we discovered a positive correlation between the number of "Major" contributors and the number of bugs, which became stronger with the new assignment model. These findings align with previous studies conducted by Bird et al. [9] and Greiler et al. [18] in `Microsoft` projects. We also observed that all the correlations became stronger after implementing the new assignment model, except for the correlation between "Modest" contributors (<50%) and code size with the number of bugs, as reported in Table 3.

At the directory level, Table 4 presents a comparison of Spearman correlation values for metrics during the two time periods. Although the correlations generally become stronger after March 2020 and with the change in the ownership model, they all remain insignificant and close to zero. This is in contrast to the analysis conducted by Greiler et al. [18], which showed much stronger correlations between these metrics and the number of bugs in `Microsoft`.

> *Our analysis of Brighsquid projects at the file level revealed a consistent negative correlation between the number of bugs and the ownership metric for both time periods. This finding is in line with state-of-the-art research in the field. Furthermore, when comparing the results before and after March 2020, we observed a strengthening of the negative correlation between the number of bugs and the ownership metric, indicating that the assignment of user stories to developers strengthen the negative correlation between the number of bugs and the ownership metric at Brightsquid.*

Furthermore, we utilized a multiple linear regression model to investigate **RQ1** and analyze the influence of each ownership metric at the file level on the number of bugs. Our objective was to assess the relationship between ownership measures while accounting for source code characteristics. Additionally, we aimed to determine if these effects remained consistent or varied across the two time periods. We only did this evaluation at the file level, as the correlations for the directory level were nearly zero. To achieve this, we constructed five statistical models. We began with the "Base" model, which incorporated code metrics churn, size, and complexity. We then iteratively added ownership metrics one by one. We presented a comparison of the variance in bugs ($R^2$) for the five statistical models during both time periods in Table 5. For both time periods, we observed that the inclusion of the Minor metric in our regression model resulted in the most significant improvement in explaining the variance in the number of bugs (18.8% before March 2020 and 4.5% after March 2020). On the other hand, the addition of the Major and Ownership metrics only led to marginal improvements in the variance and was statistically insignificant across all projects (3.9% before March 2020 and 0.3% after March 2020).

> *In Brightsquid, consistent with the findings in the literature, we observed a positive correlation between the number of minor contributors and the number of bugs. This correlation resulted in increased variance in the number of bugs when controlling for code metrics, both before and after the March 2020 period.*

When comparing the Base models before and after March 2020, we observe a significant improvement in the explanatory power of code metrics for the variance of bugs. This improvement is consistent across all projects and, on average, the code metrics can now account for 86.2% of the variance in bugs. This represents a notable enhancement of 22.3% compared to the pre-March 2020 period. Interestingly, we observed a considerable improvement in the resilience of projects after March 2020 toward ownership changes when controlling for code metrics. By changing the assignment process in March 2020, we observe that the differences between the Base model and the other four models are significantly lower (see Figure 5). When comparing the Base model with Base + Minor before March 2020, adding the Minor metric to the model increased the proportion of variance by 18.8% however, the addition of Minors accounts for only a 4.5% increase post-March 2020.

**Table 4: Spearman correlation coefficients between metrics and the number of bug fixes on directory level.**

| Ownership Metrics | Proj1 | Proj2 | Proj3 | Proj4 | Avg. |
|---|---|---|---|---|---|
| *March 2018 to March 2020* | | | | | |
| Avgownership | 0.219 | -0.121 | 0.026 | 0.136 | 0.065 |
| Ownershipdir | 0.054 | -0.124 | 0.126 | 0.136 | 0.048 |
| Minownerdir | 0.054 | -0.121 | 0.162 | 0.140 | 0.058 |
| Pcminors | 0.055 | -0.109 | 0.202 | 0.139 | 0.071 |
| Pcminimals | 0.044 | -0.156 | 0.130 | 0.125 | 0.035 |
| Pcmajors | 0.057 | -0.136 | -0.037 | 0.130 | 0.003 |
| Avgminors | 0.212 | -0.120 | 0.027 | 0.135 | 0.063 |
| Minownedfile | 0.141 | -0.052 | 0.177 | 0.171 | 0.109 |
| Avgminimals | 0.210 | -0.121 | 0.026 | 0.134 | 0.062 |
| Weakowneds | 0.355 | 0.110 | -0.060 | -0.148 | 0.064 |
| Aavgcontributors | 0.361 | 0.251 | 0.441 | 0.27 | 0.33 |
| *March 2020 to March 2022* | | | | | |
| Avgownership | -0.292 | -0.046 | 0.101 | 0.108 | -0.032 |
| Ownershipdir | -0.041 | -0.032 | 0.155 | 0.113 | 0.049 |
| Minownerdir | -0.041 | -0.038 | 0.103 | 0.115 | 0.034 |
| Pcminors | -0.041 | -0.031 | 0.105 | 0.169 | 0.05 |
| Pcminimals | -0.037 | -0.035 | 0.102 | 0.122 | 0.038 |
| Pcmajors | -0.043 | -0.035 | 0.064 | 0.167 | 0.038 |
| Avgminors | -0.299 | -0.046 | 0.101 | 0.101 | -0.035 |
| Minownedfile | -0.170 | -0.047 | 0.144 | 0.429 | 0.089 |
| Avgminimals | -0.289 | -0.055 | 0.101 | 0.154 | -0.022 |
| Weakowneds | 0.363 | 0.046 | -0.179 | -0.181 | 0.012 |
| Avgcontributors | 0.175 | 0.244 | 0.516 | 0.37 | 0.32 |

Table 5: Variance in bugs for the Base model (Code metrics) and models with Minor, Major, and Ownership added. An asterisk$^*$ denotes that a model showed models statistically significant improvement when the additional variable was added.

| Model | March 2018 to March 2020 | | | | |
|---|---|---|---|---|---|
| | Proj1 | Proj2 | Proj3 | Proj4 | Avg. |
| Base (Code metrics) | 41% | 98.1% | 31.1% | 85.5% | 63.9% |
| Base + Total | 82%* (+41.0%) | 99.7%* (+1.6%) | 70.0%* (+29.0%) | 88.9%* (+3.3%) | 85.1% (+18.7%) |
| Base + Minor | 83%* (+42.0%) | 99.7%* (+1.6%) | 70.8%* (+29.7%) | 87.6%* (+2.1%) | 85.2% (+18.8%) |
| Base + Minor + Major | 83% (+0%) | 99.7%(+0%) | 69.9%(-0.9%) | 88.6%* (+1.0%) | 85.3%(+0%) |
| Base + Minor + Major + Ownership | 83% (+0%) | 99.7% (+0%) | 77.7%* (+7.8%) | 96.7%* (+8.1%) | 89.2% (+3.9%) |
| | March 2020 to March 2022 | | | | |
| Base (Code metrics) | 79% | 90.8% | 77.6% | 97.7% | 86.2% |
| Base + Total | 80%* (+1%) | 91.0%+(+0.2%) | 93.0%* (+15.4%) | 98.9%* (+1.2%) | 90.7% (+4.4%) |
| Base + Minor | 98.9%* (+1.2%) | 91.0% (+0.2%) | 93.0%* (+15.4%) | 98.9%* (+1.2%) | 95.4% (+4.5%) |
| Base + Minor + Major | 98.9% (+0%) | 98.3%* (+7.3%) | 93.0%(+0%) | 98.9% (+0%) | 97.2% (+1.8%) |
| Base + Minor + Major + Ownership | 99.9%* (+1%) | 98.7% (+0.4%) | 93.0% (+0%) | 98.9% (+0%) | 97.6% (+0.3%) |

> *The change in the assignment model in March 2020 effectively mitigated the impact of Minor contributors on the number of bugs when controlling for code metrics. This resulted in a reduction of 14.3% in the contribution of Minor contributors to the variance.*

### 5.2 Bug predictability performance (RQ2)

In Table 6 we provided a summary and comparison of the performance metrics (precision, recall, and F-measure) of the Random Forest classifier for classifying defective sources at both the file and directory levels for Brightsquid projects.

At the file level, considering all four projects from March 2018 to March 2020 and March 2020 to March 2022, we achieved an average F-measure of 0.52 and 0.63, respectively. Moving on to the directory level, we attained an average F-measure of 0.64 and 0.70, respectively, for the same time periods. Analyzing the results presented in Table 6, it is evident that the Random Forest classifier performs better for identifying defective files and directories in the period from March 2020 to March 2022, compared to the period from March 2018 to March 2020, across all projects, with the exception of Proj2 at the file level.

> *The Random Forest model demonstrates improved precision and recall in predicting buggy files and directories following the change in the assignment model in 2020 at Brighsquid.*

In line with the approach taken by Greiler et al. [18], we conducted a metric importance analysis to assess the predictive power of individual performance metrics at both the file and directory levels. Table 7 summarize our file-level observation and Table 8 summarizes the metric importance at the directory level. However, we only focused on the files and directories that have been changed during each time period. At the file level, the average importance scores for all ownership metrics increased for the four projects during the period from March 2020 to March 2022. Notably, Minors exhibited the highest predictive power for defective files, accounting for 17.5% before March 2020 and 27.8% after March 2020.

> *After the change in the assignment model, all file ownership metrics showed an increase in predictive power in and across all the projects. The number of Minor contributors is the metric with the highest importance score for both time periods.*

Moving to the directory level, as shown in Table 8, our metric importance analysis indicated that the average percentage of commits by the highest contributor in a directory (Ownershipdir) possessed the highest average importance score (19.4%) for predicting defective source directories across all four projects from March 2018 to March 2020. During this period, the second most important metric on average was the average ownership value of a directory. However, after the process change in March 2020, the percentage of contributors with less than 50% commits (Modests) demonstrated the highest predictive power (20.9%) at the directory level. Analyzing the average importance scores of these directory-level metrics, we found that the scores of Avgownership, Pcminors, Pcmajors, Avgminors, Minownedfile, and Avgminimals increased after March 2020, while the importance scores of other metrics were higher prior to March 2020. Hence the results are inconclusive.

### 5.3 Survey with developers (RQ3)

A total of 67 developers participated in the survey, out of which 40 developers completed the survey. We discard the incomplete

Table 6: Details on Precision, Recall, and F-measure for predicting defective files.

| | File level | | | | | |
|---|---|---|---|---|---|---|
| | March 2018 to March 2020 | | | March 2020 to March 2022 | | |
| ID | Precision | Recall | F-measure | Precision | Recall | F-measure |
| Proj1 | 0.55 | 0.40 | 0.46 | 0.67 | 0.65 | 0.66 |
| Proj2 | 0.53 | 0.38 | 0.44 | 0.45 | 0.40 | 0.43 |
| Proj3 | 0.44 | 0.65 | 0.52 | 0.69 | 0.65 | 0.67 |
| Proj4 | 0.70 | 0.63 | 0.66 | 0.77 | 0.73 | 0.75 |
| Avg. | 0.55 | 0.51 | 0.52 | 0.64 | 0.61 | 0.63 |
| | Directory level | | | | | |
| Proj2 | 0.67 | 0.61 | 0.64 | 0.68 | 0.68 | 0.68 |
| Proj3 | 0.71 | 0.65 | 0.68 | 0.79 | 0.75 | 0.77 |
| Proj4 | 0.69 | 0.61 | 0.65 | 0.71 | 0.69 | 0.70 |
| Avg. | 0.69 | 0.60 | 0.64 | 0.71 | 0.69 | 0.70 |

Table 7: Metric importance in accordance to ownership metrics across the changed files

| | March 2018 to March 2020 | | | | |
|---|---|---|---|---|---|
| Projects | Ownership | Major | Minors | Minimals | Total |
| Proj1 | 0.09 | 0.13 | 0.19 | 0.21 | 0.25 |
| Proj2 | 0.13 | 0.12 | 0.23 | 0.15 | 0.11 |
| Proj3 | 0.22 | 0.13 | 0.18 | 0.12 | 0.15 |
| Proj4 | 0.11 | 0.15 | 0.17 | 0.14 | 0.15 |
| Avg. | 0.14 | 0.13 | 0.17 | 0.14 | 0.15 |
| | March 2020 to March 2022 | | | | |
| Proj1 | 0.17 | 0.16 | 0.30 | 0.10 | 0.25 |
| Proj2 | 0.20 | 0.17 | 0.23 | 0.13 | 0.28 |
| Proj3 | 0.24 | 0.21 | 0.27 | 0.27 | 0.17 |
| Proj4 | 0.23 | 0.15 | 0.32 | 0.20 | 0.40 |
| Avg. | 0.21 | 0.17 | 0.28 | 0.17 | 0.27 |

**Table 8: Metric importance in accordance with ownership metrics across the changed directories**

| Metrics | Proj1 | Proj2 | Proj3 | Proj4 | Avg. |
|---|---|---|---|---|---|
| *March 2018 to March 2020* | | | | | |
| Avgownership | 0.10 | 0.17 | 0.22 | 0.27 | 0.19 |
| Ownershipdir | 0.24 | 0.10 | 0.33 | 0.10 | 0.19 |
| Minownerdir | 0.13 | 0.21 | 0.17 | 0.16 | 0.1 |
| Pcminors | 0.19 | 0.08 | 0.18 | 0.28 | 0.19 |
| Pcminimals | 0.05 | 0.09 | 0.05 | 0.12 | 0.17 |
| Pcmajors | 0.10 | 0.12 | 0.10 | 0.17 | 0.12 |
| Avgminors | 0.08 | 0.13 | 0.09 | 0.09 | 0.10 |
| Minownedfile | 0.14 | 0.13 | 0.18 | 0.19 | 0.16 |
| Avgminimals | 0.10 | 0.09 | 0.15 | 0.09 | 0.11 |
| Weakowneds | 0.12 | 0.12 | 0.10 | 0.17 | 0.12 |
| Avgcontributors | 0.13 | 0.21 | 0.08 | 0.09 | 0.13 |
| *March 2020 to March 2022* | | | | | |
| Avgownership | 0.17 | 0.18 | 0.34 | 0.12 | 0.20 |
| Ownershipdir | 0.21 | 0.13 | 0.27 | 0.08 | 0.17 |
| Minownerdir | 0.22 | 0.20 | 0.15 | 0.13 | 0.17 |
| Pcminors | 0.21 | 0.14 | 0.19 | 0.29 | 0.21 |
| Pcminimals | 0.09 | 0.14 | 0.08 | 0.15 | 0.11 |
| Pcmajors | 0.11 | 0.08 | 0.10 | 0.15 | 0.11 |
| Avgminors | 0.10 | 0.14 | 0.10 | 0.13 | 0.12 |
| Minownedfile | 0.15 | 0.13 | 0.17 | 0.22 | 0.17 |
| Avgminimals | 0.09 | 0.15 | 0.10 | 0.12 | 0.11 |
| Weakowneds | 0.10 | 0.13 | 0.09 | 0.09 | 0.10 |
| Avgcontributors | 0.09 | 0.19 | 0.08 | 0.07 | 0.11 |

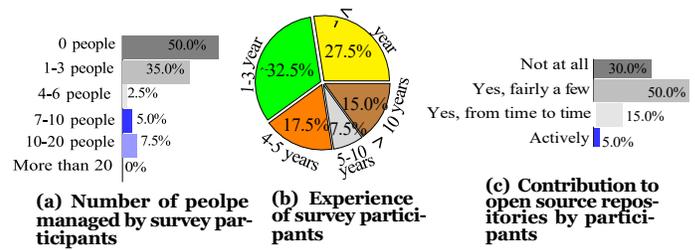

(a) Number of people managed by survey participants

(b) Experience of survey participants

(c) Contribution to open source repositories by participants

**Figure 3: Demographics of survey participants**

responses and present the results of the complete survey. 38 of the participants had the title of software developer in their team. The survey consisted of three demographic questions, the results of which are presented in Figure 3. 32.5% of the participants had one to three years of experience, 27.5% had less than one year of experience, and 15.0% had more than 10 years of experience. Regarding team management, 50.0% of the participants did not manage any developers in their team while 7.5% managed 10-20 developers. 70% of our participants were contributing to open-source projects.

Table 9 summarizes the questions and results regarding the source and degree of developers' sense of ownership. 85.0% of our participants expressed that they feel ownership toward the software artifacts. Among them, 20.95% felt most ownership toward code, 14.86% toward Tasks, and 13.51% toward bugs or issues. When asked about the circumstances in which they experience the strongest sense of ownership for an artifact, 35% of our participants stated that it occurs when they are the sole authors of the artifact, while 22.5% mentioned feeling a heightened sense of ownership when they contribute more than others. Additionally, 17.5% of the developers reported feeling ownership based on their higher knowledge and expertise, while 12.5% attributed their sense of ownership to their historical contributions in maintaining the artifact. However, 60.0% of them indicated that they frequently or always perceive the ownership of artifacts as shared between themselves and others.

> *Developers mainly perceive ownership towards code and tasks, and they primarily determine their ownership based on their level of contribution, which accounts for the majority of cases (57.5%).*

We provided the developers with a dictionary definition of accountability, which states that "Accountability refers to the real or perceived likelihood that the actions, decisions, or behaviors of an individual, group, or organization will be evaluated. It also entails the potential for the individual, group, or organization to receive rewards or sanctions based on this expected evaluation." After presenting this definition, we proceeded to ask the participants to identify the person towards whom they feel the highest level of accountability. Among developers, team leads and managers hold the highest level of accountability (84.2% of participants). Following that, developers feel accountable towards themselves (79%) and the developers express accountability towards their teammates and co-workers (73%) in the third place.

We conducted a survey in which we presented a series of 13 statements regarding accountability within their current software team. The developers were asked to indicate their level of agreement with each statement using a five-point Likert scale. The results, shown in Figure 4, revealed interesting insights.

A significant majority of participants, 94.7%, expressed agreement in feeling accountable for the quality of the outcome of the tasks assigned to them. Similarly, 92.1% of developers felt accountable for the procedure and steps taken to fulfill the assigned tasks. When it came to the tasks specifically assigned to them, 92.1% felt accountable without any disagreement. However, for tasks they volunteered to take, slightly fewer developers, 81.5%, expressed accountability with no disagreement. Moreover, 89.5% of developers showed a sense of accountability for timely task delivery. Interestingly, in the scenario where a teammate is unable to maintain an artifact developed by the participating developer, 44.8% of participants disagreed with feeling accountable for it, indicating a lower level of accountability in this particular situation.

**Table 9: Source and degree of developer's sense of ownership**

| Do you feel ownership towards any software artifacts? | % of Participants |
|---|---|
| Yes | 85.0% |
| No | 15.0% |
| **Which artifact do you feel most ownership towards?** | **% of Participants** |
| Code | 20.95% |
| Task | 14.86% |
| Bug/Issue | 13.51% |
| Product | 9.46% |
| Test | 9.46% |
| User Story | 9.46% |
| Project | 8.78% |
| Requirement | 6.76% |
| Build | 5.41% |
| Others | 1.35% |
| **When do you consider yourself an artifact owner?** | **% of Participants** |
| All the artifact is written by me | 35.0% |
| I contributed more than others | 22.5% |
| I have the most knowledge and expertise on artifact | 17.5% |
| I was assigned to maintain it | 12.5% |
| Proposed the idea behind artifact (IP) | 7.5% |
| Initiated implementation | 5.0% |
| **How often do you feel ownership of an artifact is shared between you and others?** | **% of Participants** |
| Rarely | 7.50% |
| Sometimes | 32.50% |
| Frequently | 42.50% |
| Always | 17.50% |

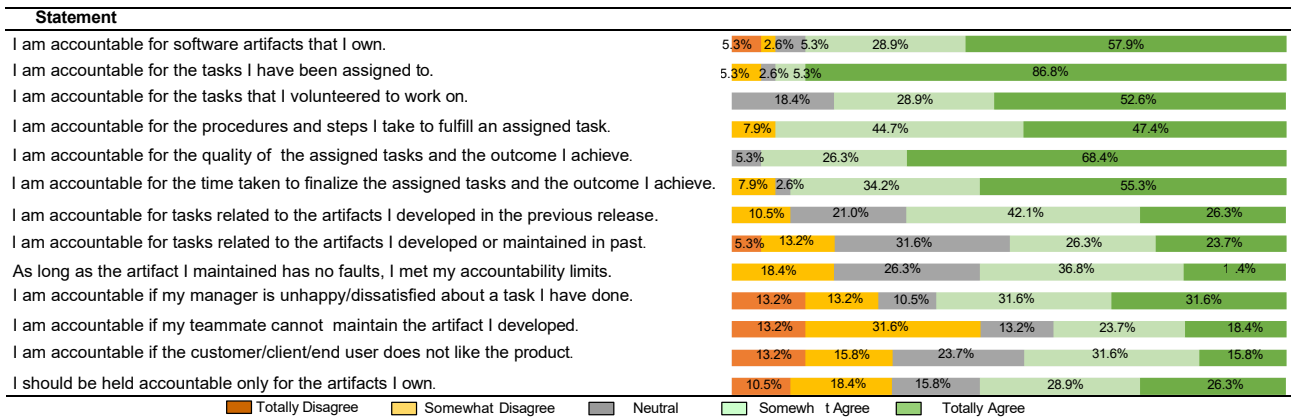

Figure 4: Accountability Levels of Developers in a Team. Participants' accountability was measured using a Likert scale.

> Developers generally agreed on being accountable for the tasks assigned to them, as well as the time and steps taken to deliver them, and the quality of the outcomes.

Only 86.8% of developers feel accountable towards the artifacts they own. We then asked the developers to rank the individuals who should be held accountable for maintaining an artifact with shared ownership. The results of their ranking are presented in Figure 5. The majority, 52.5%, believe that developers with the highest knowledge and skills should be held accountable for maintaining a shared artifact. Additionally, 62.5% of the participants believe that the person who made the last changes to the file should be accountable for its maintenance, followed by 57.5% who believe that the developer with the highest number of bug fixes in a file should be held accountable for maintaining it.

When directly asked about the relationship between accountability and ownership (Figure 6), 5% of participants considered it a weak relationship (score of 2), while 67.5% considered it a strong or very strong relationship.

> The majority of developers feel accountable towards their owned artifacts and perceive a significant relationship between accountability and ownership of these artifacts.

## 6 DISCUSSION

Two years after transitioning from task assignments to assigning user stories to developers, the leadership team at Brightsquid has observed a positive impact on developers' accountability, transforming the team culture from being solely focused on order-taking to fostering a culture of design and innovation:

*"We have to get the team more involved in the kind of design, conversation leading to the solution that we want. In the current digital health ecosystem, we do not always have a defined product for a specific market. Therefore, our continuous effort is to encourage the market to adopt new solutions. This changes our role from development management to vision management."*

While this change was inevitably driven by the increasing demand for digital health during the COVID-19 pandemic, the company has continued this practice due to the perceived higher level of accountability. Furthermore, our measurements have indicated improved code ownership across projects.

### 6.1 Assignment process and relation to accountability, ownership, and code quality

The sociotechnical aspects of software systems are inherently complex, and they contain intricate details that may have been overlooked in empirical investigations, making it challenging to establish causality. However, for this analysis, we have carefully selected four Brightsquid projects. The key distinguishing factor between the periods of 2018-2020 and 2020-2022 in these projects is the assignment model. Our selection is based on observations and has been agreed upon by the production team. Our investigation in **RQ1** and **RQ2** revealed that the change in the assignment process strengthened the correlation between ownership metrics and code quality factors. It also made code defects more predictable by using ownership metrics. These findings align with previous literature on the subject.

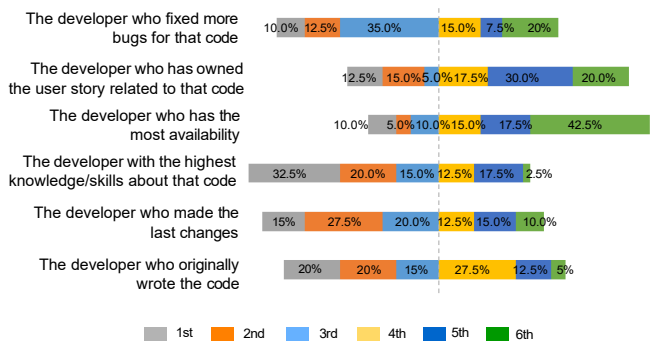

Figure 5: Who should be held accountable for maintaining an artifact with shared ownership?

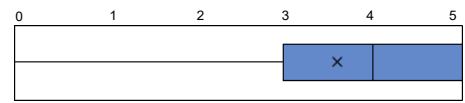

Figure 6: Developers' perceived degree of relationship between accountability and ownership (5 = strongest degree).

Our analysis at the directory level did not provide much insight into the changes when compared between the two time periods. However, through our analysis of Brighsquid projects at the file level, we discovered a consistent negative correlation between the number of bugs and the ownership metric for both time periods. This finding aligns with current research in the field. Additionally, when comparing the results before and after March 2020, we observed a strengthening of the negative correlation between the number of bugs and the ownership metric. When discussing this matter with the production team, it was perceived that the assignment of user stories to developers enhances their sense of accountability. This, in turn, ensures that the quality of the designed solution meets expectations, thereby reinforcing the negative correlation between bugs and the ownership metric at Brightsquid. In Brightsquid, the product manager does not specifically monitor ownership metrics, which include historical data on how developers have interacted with and modified a file. However, they interpret these findings as evidence that improved code quality and enhanced ownership are direct outcomes of the change in the assignment process. This change involves assigning user stories with greater delegation to the developers, making the developer responsible for the overall outcome of the story rather than just a piece of code:

*"Code quality, and I am not just talking about testing, but also considering acceptance criteria and peer reviews, depends upon the diligence of the process. The combination of people and process is what gives you the greatest likelihood of quality. Code ownership can vary on a sliding scale, and in my experience, the relevance of how many people touch the code becomes less significant when the process is well-defined and delegated appropriately."*

We observed a stronger correlation and predictive power between ownership metrics and the number of bugs at the file level (Table 3 and Table 5). Interestingly, when comparing the variance in the number of bugs controlled by code metrics (Table 5), we found:

**First,** the base model, which measures variance in the number of bugs based on code metrics, experienced a significant increase of 38% in enhanced variance. This indicates that code attributes alone can explain a considerable degree of variance in the number of bugs after the change in March 2020.

**Second,** the quality of all projects demonstrates greater resilience to change and ownership metrics. Previously, the addition of minor contributors could improve the variance in bugs by 18.8% compared to the base model. However, following the process change in 2020, this improvement was reduced to 4.5%.

We interpret this as the shift in the assignment model distributing ownership among all team members and shifting the mindset from code perfection to implementing feature sets and releasing the product. Consequently, the significance of who owns the code and their historical contributions in a file becomes less relevant in defining its quality.

*"... along with engaging developers more in design they became more diligent about releasing the product and less about technical quality (to balance their efforts). This helped in reducing risk and adapting to the change during the pandemic because we could see if the market adapts to what we have built."*

With the change in the assignment model, the team spends less effort on pre-defining the requirements which lead into identifying more requirement-related bugs after development;

*"We do less diligence testing before a release hence the bugs can go up. We spent way less time on requirements hence there were more issues coming up because the features are less specified in advance. We became more reactive toward bugs and improvement requests."*

## 6.2 Survey Implications (RQ3)

When surveying the developers, it was found that 20.95% felt ownership towards code, while 14.86% felt ownership towards tasks. In comparison, only 9.46% perceived ownership towards user stories, and 6.76% towards requirements. Furthermore, when asked about accountability, the majority (52.7%) expressed controversial or neutral feelings towards being accountable for the satisfaction of customers or end users. This is whilst the survey showed a clear tendency of developers to feel accountable toward assigned tasks. When discussing these results with Brighsquid, the product manager stated:

*"It all depends on the team culture. In Brightsquid, it is time to time our team asking that what's the task but usually we go around that by starting with a story, gathering the information like investigating the story and presenting all the facts about what is happening, and then we talk with the team on that story to decide the designer solution going forward."*

We observed a strong correlation of 0.82 between the participation of users in open-source development and their accountability toward end-user satisfaction. This suggests that the open-source mindset fosters a culture that encompasses a broader spectrum of accountability. In addition, we employed the Mann-Whitney test to assess potential differences in group means for the sense of accountability in a team (Figure 4) among open-source participants. Our analysis revealed that users with experience in open-source development feel significantly more accountable if their teammates cannot maintain the artifact they have developed ($p-value = 0.008$).

## 7 THREATS TO VALIDITY

This study should be interpreted within the limitations of being a case study for research questions **RQ1** and **RQ2** [26, 47]. Our focus was solely on the four most active projects within Brightsquid, which may restrict the generalizability of our findings to other organizations or open-source projects. However, it provides a fair evaluation of the situation within Brightsquid, as the majority of developers and resources are involved in these projects. We carefully observed all relevant factors before and after March 2020 to identify the roots of ownership changes, and no significant changes other than the assignment model were found. Nonetheless, as a sociotechnical system, the software team is complex, and there may be confounding factors that we did not identify. In our case study, we considered ownership metrics used in previous studies [9, 18], and it's worth noting that using different ownership metrics may yield different findings. We acknowledge a limitation of our study, as we focused solely on individual ownership levels and did not consider team ownership within Brightsquid. This limitation can impact the internal validity of our findings. To address this gap, future studies should include an examination of team ownership levels, as this may provide a more comprehensive understanding of how ownership impacts software development dynamics.

For the survey of **RQ3**, the conclusions drawn in our study are based on a survey, which may introduce some inaccuracies and the number of participants is limited, the larger group of participants can make our conclusion more vigorous. Surveying software developers may not always capture a comprehensive perspective of real-world practices [14, 26]. However, surveys have been widely used as a research tool in software engineering. We view the survey as a complementary approach to other types of studies, such as mining, in order to gain a more holistic understanding of the research topic.

## 8 RELATED WORK

While accountability has been discussed in various teams as a scientific discipline [25, 29], it has not been specifically studied within software teams. In this context, we provide a concise overview of related studies on accountability and ownership in software teams.

### 8.1 Individual and team accountability

Software development is a collaborative endeavor, and its social aspects have long been a topic of discussion. Many parallels can be drawn between the social behavior of human society and various aspects of software development. Research has explored the relationship between psychological ownership, self-identity, organizational accountability, a sense of belonging, and organizational citizenship [10]. Interestingly, psychological ownership has received limited attention in the field of software engineering [11, 12].

A sense of ownership inherently encompasses a sense of responsibility towards a target. This responsibility entails protecting and improving possessions and may involve controlling or limiting access by others [5]. On the other hand, accountability involves the expected rights and responsibilities of individuals, including the right to hold others accountable and the expectation of being held accountable oneself [13]. Feelings of psychological ownership are closely tied to attachment towards places, objects, and people [5, 13]. The need for belongingness in the organization or workplace [42] is fulfilled when individuals feel like owners within the organization, satisfying their socio-emotional needs [20, 36].

### 8.2 Ownership in Software Teams

Ownership within software teams plays a pivotal role in fostering commitment, initiative, and delivering high-quality work [9, 21, 28, 33, 34, 41]. It encompasses two primary forms: Psychological Ownership, which relates to an individual's sense of possession [12], and Corporal Ownership, which pertains to the developmental history of software artifacts. Code ownership is widely recognized and facilitates accountability, task delegation, and identification of experts within a software product [39]. Research indicates that weak code ownership is associated with an increased likelihood of defects [8, 9, 17, 18]. In the agile context, task ownership is particularly important as it empowers team members to take responsibility for assigned tasks from task boards, thereby enhancing collaboration and project progress [43]. Bug ownership serves as a means to identify the responsible developers involved in bug-fixing efforts, whether it be the developer who introduced the code or another team member [48]. Additionally, test ownership designates individuals who contribute to specific tests within a test suite, streamlining test management and accountability [22]. Ownership and task assignment are closely interconnected [16, 17, 19]. Typically, the team lead assumes responsibility for task and bug assignments, investing significant time in determining the most suitable developer for each task [2]. Incorporating ownership considerations in task assignment enhances the efficiency and effectiveness of bug resolution within the team [1, 3, 6, 46]. Several systematic literature reviews have provided comprehensive insights into the relationship between code ownership, quality, and task assignments [4, 27, 35].

## 9 CONCLUSION

Brighsquid implemented a new task assignment process aimed at enhancing developer accountability toward customers. The findings from both code change analysis and a survey of 67 developers demonstrate the positive impact of increased accountability on code ownership. Notably, the revised assignment model not only heightened the perceived sense of accountability within the production team but also improved the resilience of the code against ownership changes. Moreover, when comparing the data before and after March 2020, there was a notable reinforcement of the negative correlation between the number of bugs and the ownership metric. This observation suggests that the assignment of user stories played a significant role in reducing the occurrence of bugs. The survey results revealed an intriguing statistic, with 67.5% of developers associating perceived accountability with artifact ownership. This empirical examination sheds light on accountability within software teams, opening up new avenues for further evaluation. By understanding the impact of accountability on ownership, future research can focus on developing effective strategies to foster motivation and a sense of belonging within teams, ultimately leading to increased productivity.

## REFERENCES
[1] J. Anvik. Automating bug report assignment. In *Proceedings of the 28th international conference on Software engineering*, pages 937–940, 2006.
[2] J. Anvik, L. Hiew, and G. C. Murphy. Who should fix this bug? In *Proceedings of the 28th international conference on Software engineering*, pages 361–370, 2006.
[3] J. Anvik and G. C. Murphy. Determining implementation expertise from bug reports. In *Fourth International Workshop on Mining Software Repositories (MSR'07: ICSE Workshops 2007)*, pages 2–2. IEEE, 2007.
[4] A. Arunarani, D. Manjula, and V. Sugumaran. Task scheduling techniques in cloud computing: A literature survey. *Future Generation Computer Systems*, 91:407–415, 2019.
[5] J. B. Avey, B. J. Avolio, C. D. Crossley, and F. Luthans. Psychological ownership: Theoretical extensions, measurement and relation to work outcomes. *Journal of Organizational Behavior: The International Journal of Industrial, Occupational and Organizational Psychology and Behavior*, 30(2):173–191, 2009.
[6] O. Baysal, M. W. Godfrey, and R. Cohen. A bug you like: A framework for automated assignment of bugs. In *2009 IEEE 17th International Conference on*

*Program Comprehension*, pages 297–298. IEEE, 2009.

[7] C. I. Bezerra, J. C. de Souza Filho, E. F. Coutinho, A. Gama, A. L. Ferreira, G. L. de Andrade, and C. E. Feitosa. How human and organizational factors influence software teams productivity in covid-19 pandemic: A brazilian survey. In *Proceedings of the XXXIV Brazilian Symposium on Software Engineering*, pages 606–615, 2020.

[8] C. Bird, N. Nagappan, P. Devanbu, H. Gall, and B. Murphy. Does distributed development affect software quality? an empirical case study of windows vista. *Communications of the ACM*, 52(8):85–93, 2009.

[9] C. Bird, N. Nagappan, B. Murphy, H. Gall, and P. Devanbu. Don't touch my code! examining the effects of ownership on software quality. In *Proceedings of the 19th ACM SIGSOFT symposium and the 13th European conference on Foundations of software engineering*, pages 4–14, 2011.

[10] I. Buchem. Psychological ownership and personal learning environments: Do sense of ownership and control really matter. In *PLE Conference Proceedings*, volume 1. Citeseer, 2012.

[11] T. Chung, P. Sharma, and D. Sherae. The impact of person-organization fit and psychological ownership on turnover in open source software projects. 2015.

[12] T. R. Chung, P. N. Sharma, and S. L. Daniel. The impact of person-organization fit and psychological ownership on turnover in open source software projects. In *AMCIS*, 2015.

[13] K. S. Corts. Teams versus individual accountability: solving multitask problems through job design. *The RAND Journal of Economics*, 38(2):467–479, 2007.

[14] T. Dyba, B. A. Kitchenham, and M. Jorgensen. Evidence-based software engineering for practitioners. *IEEE software*, 22(1):58–65, 2005.

[15] D. Ford, M.-A. Storey, T. Zimmermann, C. Bird, S. Jaffe, C. Maddila, J. L. Butler, B. Houck, and N. Nagappan. A tale of two cities: Software developers working from home during the covid-19 pandemic. *ACM Transactions on Software Engineering and Methodology (TOSEM)*, 31(2):1–37, 2021.

[16] M. Foucault, J.-R. Falleri, and X. Blanc. Code ownership in open-source software. In *Proceedings of the 18th International Conference on Evaluation and Assessment in Software Engineering*, pages 1–9, 2014.

[17] M. Foucault, C. Teyton, D. Lo, X. Blanc, and J.-R. Falleri. On the usefulness of ownership metrics in open-source software projects. *Information and Software Technology*, 64:102–112, 2015.

[18] M. Greiler, K. Herzig, and J. Czerwonka. Code ownership and software quality: A replication study. In *2015 IEEE/ACM 12th Working Conference on Mining Software Repositories*, pages 2–12. IEEE, 2015.

[19] P. J. Guo, T. Zimmermann, N. Nagappan, and B. Murphy. " not my bug!" and other reasons for software bug report reassignments. In *Proceedings of the ACM 2011 conference on Computer supported cooperative work*, pages 395–404, 2011.

[20] A. T. Hall, M. G. Bowen, G. R. Ferris, M. T. Royle, and D. E. Fitzgibbons. The accountability lens: A new way to view management issues. *Business Horizons*, 50(5):405–413, 2007.

[21] L. Hattori and M. Lanza. Mining the history of synchronous changes to refine code ownership. In *2009 6th ieee international working conference on mining software repositories*, pages 141–150. IEEE, 2009.

[22] K. Herzig and N. Nagappan. The impact of test ownership and team structure on the reliability and effectiveness of quality test runs. In *Proceedings of the 8th ACM/IEEE International Symposium on Empirical Software Engineering and Measurement*, pages 1–10, 2014.

[23] R. Hoda and L. K. Murugesan. Multi-level agile project management challenges: A self-organizing team perspective. *Journal of Systems and Software*, 117:245–257, 2016.

[24] R. Hoda and J. Noble. Becoming agile: a grounded theory of agile transitions in practice. In *2017 IEEE/ACM 39th International Conference on Software Engineering (ICSE)*, pages 141–151. IEEE, 2017.

[25] J. R. Katzenbach and D. K. Smith. The discipline of teams. *Harvard business review*, 83(7):162, 2005.

[26] D. Lo, N. Nagappan, and T. Zimmermann. How practitioners perceive the relevance of software engineering research. In *Proceedings of the 2015 10th Joint Meeting on Foundations of Software Engineering*, pages 415–425, 2015.

[27] S. Mahmood, S. Anwer, M. Niazi, M. Alshayeb, and I. Richardson. Key factors that influence task allocation in global software development. *Information and Software Technology*, 91:102–122, 2017.

[28] L. M. Maruping, X. Zhang, and V. Venkatesh. Role of collective ownership and coding standards in coordinating expertise in software project teams. *European Journal of Information Systems*, 18:355–371, 2009.

[29] L. M. Marx and F. Squintani. Individual accountability in teams. *Journal of Economic Behavior & Organization*, 72(1):260–273, 2009.

[30] C. Miller, P. Rodeghero, M.-A. Storey, D. Ford, and T. Zimmermann. " how was your weekend?" software development teams working from home during covid-19. In *2021 IEEE/ACM 43rd International Conference on Software Engineering (ICSE)*, pages 624–636. IEEE, 2021.

[31] M. Nayebi, L. Dicke, R. Ittyipe, C. Carlson, and G. Ruhe. Essmart way to manage customer requests. *Empirical Software Engineering*, 24:3755–3789, 2019.

[32] M. Nayebi, S. J. Kabeer, G. Ruhe, C. Carlson, and F. Chew. Hybrid labels are the new measure! *IEEE Software*, 35(1):54–57, 2017.

[33] M. Nayebi, K. Kuznetsov, P. Chen, A. Zeller, and G. Ruhe. Anatomy of functionality deletion: an exploratory study on mobile apps. In *Proceedings of the 15th International Conference on Mining Software Repositories*, pages 243–253, 2018.

[34] M. Nayebi, G. Ruhe, and T. Zimmermann. Mining treatment-outcome constructs from sequential software engineering data. *IEEE Transactions on Software Engineering*, 47(2):393–411, 2019.

[35] A. Niknafs, J. Denzinger, and G. Ruhe. A systematic literature review of the personnel assignment problem. In *Proceedings of the International Multiconference of Engineers and Computer Scientists, Hong Kong*, 2013.

[36] H. Nissenbaum. Computing and accountability. *Communications of the ACM*, 37(1):72–81, 1994.

[37] M. E. Nordberg. Managing code ownership. *IEEE software*, 20(2):26–33, 2003.

[38] S. L. Pfleeger and B. A. Kitchenham. Principles of survey research: part 1: turning lemons into lemonade. *ACM SIGSOFT Software Engineering Notes*, 26(6):16–18, 2001.

[39] F. Rahman and P. Devanbu. Ownership, experience and defects: a fine-grained study of authorship. In *Proceedings of the 33rd International Conference on Software Engineering*, pages 491–500, 2011.

[40] D. Russo, P. H. Hanel, S. Altnickel, and N. Van Berkel. The daily life of software engineers during the covid-19 pandemic. In *2021 IEEE/ACM 43rd International Conference on Software Engineering: Software Engineering in Practice (ICSE-SEIP)*, pages 364–373. IEEE, 2021.

[41] T. Sedano, P. Ralph, and C. Péraire. Practice and perception of team code ownership. In *Proceedings of the 20th international conference on evaluation and assessment in software engineering*, pages 1–6, 2016.

[42] J. Shakespeare-Finch and E. Daley. Workplace belongingness, distress, and resilience in emergency service workers. *Psychological Trauma: Theory, Research, Practice, and Policy*, 9(1):32, 2017.

[43] Y. Shastri, R. Hoda, and R. Amor. Understanding the roles of the manager in agile project management. In *Proceedings of the 10th Innovations in Software Engineering Conference*, pages 45–55, 2017.

[44] A. Taivalsaari, T. Mikkonen, and K. Systä. Liquid software manifesto: The era of multiple device ownership and its implications for software architecture. In *2014 IEEE 38th Annual Computer Software and Applications Conference*, pages 338–343. IEEE, 2014.

[45] P. Thongtanunam, S. McIntosh, A. E. Hassan, and H. Iida. Revisiting code ownership and its relationship with software quality in the scope of modern code review. In *Proceedings of the 38th international conference on software engineering*, pages 1039–1050, 2016.

[46] C. Weiss, R. Premraj, T. Zimmermann, and A. Zeller. How long will it take to fix this bug? In *Fourth International Workshop on Mining Software Repositories (MSR'07: ICSE Workshops 2007)*, pages 1–1. IEEE, 2007.

[47] C. Wohlin, P. Runeson, M. Höst, M. C. Ohlsson, B. Regnell, and A. Wesslén. *Experimentation in software engineering*. Springer Science & Business Media, 2012.

[48] W. Zhu and M. W. Godfrey. Mea culpa: How developers fix their own simple bugs differently from other developers. In *2021 IEEE/ACM 18th International Conference on Mining Software Repositories (MSR)*, pages 515–519. IEEE, 2021.